\documentclass[aps,pra,a4paper,twocolumn,noshowpacs,superscriptaddress,floatfix]{revtex4}
\usepackage{amsfonts}
\usepackage{amsmath}
\usepackage{amssymb}
\usepackage{array}
\usepackage{graphicx}
\usepackage{epstopdf}
\usepackage{multirow}
\usepackage{booktabs}
\usepackage{makecell}
\usepackage{setspace}
\usepackage{booktabs}
\usepackage{threeparttable}

\newcommand{\PreserveBackslash}[1]{\let\temp=\\#1\let\\=\temp}
\newcolumntype{C}[1]{>{\PreserveBackslash\centering}p{#1}}
\newcolumntype{R}[1]{>{\PreserveBackslash\raggedleft}p{#1}}
\newcolumntype{L}[1]{>{\PreserveBackslash\raggedright}p{#1}}

\begin{document}

\title{Application of the Hylleraas-$B$-spline basis set: Nonrelativistic Bethe logarithm of helium}

\author{San-Jiang Yang}
\affiliation{Department of Physics, Wuhan University, Wuhan 430072, China}
\affiliation{State Key Laboratory of Magnetic Resonance and Atomic and Molecular Physics, Wuhan Institute of Physics and Mathematics,
Chinese Academy of Sciences, Wuhan 430071, Peoples Republic of China}

\author{Yong-Bo Tang}
\affiliation{College of Engineering Physics, Shenzhen technology University, Shenzhen 518118, China}

\author{Yong-Hua Zhao}
\affiliation{Computer Network Information Center, Chinese Academy of Sciences Beijing, China}

\author{Ting-Yun Shi \footnote{electronic mail: tyshi@wipm.ac.cn}}
\affiliation{State Key Laboratory of Magnetic Resonance and Atomic and Molecular Physics, Wuhan Institute of Physics and Mathematics,
Chinese Academy of Sciences, Wuhan 430071, Peoples Republic of China}

\author{Hao-Xue Qiao \footnote{electronic mail: qhx@whu.edu.cn}}
\affiliation{Department of Physics, Wuhan University, Wuhan 430072, China}

\begin{abstract}
In this work, we report an application of Hylleraas-$B$-spline basis set to the nonrelativistic Bethe logarithm calculation of helium.  The Bethe logarithm for $n\ ^1S$, $n$ up to 10, states of helium are calculated with a precision of 7-9 significant digits in two gauges, which greatly improves the accuracy of the traditional $B$-spline basis set. In addition, to deal with the numerical linear correlation problem in Bethe logarithm calculation, we developed a multiple-precision generalized symmetric eigenvalue problem solver (MGSEPS). This program may be very useful to precision calculations.

\end{abstract}
\maketitle

\section{INTRODUCTION}
Driven by precision measurements of few body atoms and molecules, combining high accuracy experimental data and theoretical calculations to extract the basic physical constants enters a new and more arduous stage~\cite{pachucki2017testing,PhysRevLett.92.023001,zheng2017x,RevModPhys.90.025008}. Among few body systems, helium has attracted the attention of both theoretical and experimental researchers for a long time. Compared to hydrogen, the $2\; ^3P$ state of helium not only has a longer lifetime but has a wider interval of the fine structure splitting. Since Schwartz first proposed in 1964, helium is recognized as the best atom system to extract the value of fine-structure constant $\alpha$ up to now~\cite{schwartz1964fine,zheng2017x}. On the other hand, the high precise measurements for $2 \; ^3S \rightarrow 2 \; ^1S$ and $2 \; ^3S \rightarrow 2 \; ^3P$ of helium can be used to determine the nuclear charge radius, which may help to solve the proton size puzzle~\cite{PhysRevA.98.040501,Leeuwen_2006,PhysRevLett.74.3553,PhysRevA.73.034502,PhysRevLett.92.023001,PhysRevLett.108.143001}.

In the aspect of theoretical calculations, few body systems are relatively simple, which makes high precision calculation on these systems is possible~\cite{drake1978angular,yan1994general,frolov2000high-precision}. High precision calculation requires more comprehensive consideration on these systems. For the light bounded systems, nonrelativistic quantum electrodynamics (NRQED) provides an effective approach to describe their states ~\cite{caswell1986effective,jentschura2005nonrelativistic,pachucki2010fine,pachucki2017testing}. In this approach, the energy levels of a given system are expanded by fine-structure constant $\alpha$, started from the eigenvalues of the nonrelativistic hamiltonian. Except for the leading term, the other terms are regarded as various physical effect contributions to nonrelativistic energy levels. In these terms, the Bethe logarithm (BL) is particularly remarked. Generally speaking, BL type terms are caused by the contributions of low-energy virtual photons, which makes the logarithmic term, $\ln|H_0-E_0|$, appears. Inserted by complete intermediate states, the BL type terms can be calculated normally. However, for precision calculation, the specific form of BL type term needs a huge interval of intermediate energy levels. This was a challenge to variational method until 1999. In that year, Drake and Goldman suggested the sum over pseudostates approach~\cite{drake1999bethe,goldman2000high}. This approach clears the obstacle and they got the numerical results of BL in about 9-10 significant digits for $n\; ^{1,3}S$ and $n\; ^{1,3}P$, $n$ up to 5, of helium-like systems. For the higher principal quantum number states, Drake also gave the $1/n$ expansions formulas in article~\cite{drake2001qed}. Laterly, there are many studies on the calculation of BL to extend the range of application of pesudostates approach to the other basis sets or systems~\cite{PhysRevA.61.052513,PhysRevA.69.054501,tang2013bethe-logarithm,PhysRevA.88.052520}. In the same year, V.I. Korobov and S.V. Korobov presented another approach which is similar to Schwartz's~\cite{PhysRev.123.1700} and V.I. Korobov presented this method in a more explicit and general form in 2012~\cite{PhysRevA.85.042514}. In Ref.~\cite{PhysRevA.85.042514}, V.I. Kovobov obtained the values of BL for the ground state of helium and $H^{+}_2$ in about 12 and 10 significant digits, respectively. At very recently, using this method, V.I. Kovobov obtained the values of BL for a wide range of principal quantum number and orbital angular momentum states of helium~\cite{KorobovBL}.

Recently, we constructed the Hylleraas-$B$-spline (H-$B$-spline) basis set and successfully applied it to the calculations of static dipole polarizabilities, dynamic dipole polarizabilities and dynamic hyperpolarizabilities of helium~\cite{yang2017application,PhysRevA.98.040501}. This basis overcomes the ground state difficulty of using the traditional $B$-spline basis and inherits the property of fitting a wider range of initial states in one diagonalization. Our previous works demonstrated the well coupling of pesudostates approach and H-$B$-spline basis. In this work, we extend this basis to the nonrelativistic Bethe logarithm of helium. Previously, Tang \textit{et al.} reported an successfully application of $B$-spline basis set to BL of hydrogen~\cite{tang2013bethe-logarithm}. And they also pointed out the close relationship between the first non-zero knot of $B$-spline and the interval of intermediate energy levels generated by $B$-spline basis. However, in the case of helium, the knots sequence of $B$-splines that are very close to the origin results in the numerical linear correlation problem. To overcome this problem, based on ARPREC~\cite{bailey2002arprec}, we developed a MPI parallel program, MGSEPS, which
can solve generalized symmetric eigenproblem effciently. With this program, we calculate the BL for $n \; ^1S$, $n$ up to 10, states of helium in acceleration gauge and velocity-acceleration gauge respectively. These results remedy the recent results of $B$-spline basis for $^1 S$ states of helium~\cite{zhang2019calculations}, and provide some comparisons for the other methods.

The article is organized as follows. In Sec. II we briefly introduce H-$B$-spline and give basic formulas of BL in two gauges. Numerical results are presented in Section III, together with a comparison with available theoretical
results. Finally, a summary is given in Sec. IV.  Atomic units (a.u.) are used throughout,
unless otherwise stated.

\section{THEORETICAL METHOD}
\subsection{Hylleraas-$B$-spline basis set}
The Hamiltonian in infinite mass approximation of helium is 
\begin{equation}
     H=\sum^{2}_{i} \left( -\frac{1}{2m} \vec \nabla_i^2 -\frac{Z}{r_i} \right) + \frac{1}{r_{12}}\ .
\end{equation}
Where $Z$ is the charge of nucleus, $m$ is the electron mass, $r_i$ is the distance between the i-th electron and nucleus, $r_{12}$ is the distance of two electrons. Considering the wave function behavior at two electron coalescences, the wave function of helium with total angular momentum $L$ and magnetic quantum number $M$ could be expanded by
\begin{equation}
\begin{aligned}
\{ \Phi _{ijcl_1l_2} = B_{i,k}(r_1)&B_{j,k}(r_2)r_{12}^{c} \\ 
&\mathbf{\Lambda}_{l_1l_2}^{LM} (\hat{r}_1 ,\hat{r}_2)  \pm exchange \}  .
\end{aligned}
\end{equation}
Where $\mathbf{\Lambda}_{l_1l_2}^{LM}$ is the vector coupled product of angular momenta $l_1$, $l_2$ for the two electrons, $B_{i,k}(r)$ represents the i-th $B$-spline functions with k order defined in a finite domain $(0,r_{max})$~\cite{deboor1978a}.
The shape of $B_{i,k}(r)$ depends on the nondecreasing knot sequence $\{ t_i \}$ (see below) and the spline order $k$. In the following calculations, the knots sequence are arranged as below:
\begin{equation}
\left\{ \begin{aligned}
           t_i&=0 \qquad \qquad \qquad \qquad \quad   i=1,2,\ldots ,k-1,\\
           t_i&=r_{max}  \frac{e^{\gamma \left( \frac{i-k}{N-k+1} \right)  }-1}{e^{\gamma}-1} \quad i=k,k+1,\ldots ,N,\\
           t_i&=r_{max} \qquad \qquad \qquad \quad \ i=N+1,\ldots ,N+k-1.
        \end{aligned}
 \right.
\end{equation}
And we set $k=7$ and $\gamma =\tau \times r_{max} $.

In this article, we restrict $c$ to less than 2. The parameters $i,j,c,l_1,l_2$ are arranged as bellow,
\begin{equation}
\begin{aligned}
           i&=1,2,\ldots ,j ,  \\
           j&=1,2,\ldots ,N  , \\
           c&=0,1 , \\
           l_1&=0,1,\ldots ,l_{max} , \\
           l_2&=0,1,\ldots ,l_{max} .
        \end{aligned}
\end{equation}
Where $N$ is called the total number of $B$-spline and $l_{max}$ is called the partial-wave expansion length. And the terms which make the norm of $\Phi _{ijcl_1l_2}$ to be zero must be eliminated. 

\subsection{Nonrelativistic Bethe logarithm}
The definition of BL in acceleration gauge as below:
\begin{equation}
\beta(n,L,S) =\frac{\mathcal{N}^{(A)}(n,L,S)}{\mathcal{D}^{(A)}(n,L,S)} \; ,
\end{equation}
where the $n,L,S$ are the quantum numbers of initial state. $\mathcal{N}^{(A)}$ and $\mathcal{D}^{(A)}$ denote numerator and denominator, the expressions are
\begin{equation}
\begin{aligned}
\mathcal{N}^{(A)}(n,L,S) =& \sum_{m,i} \left| \left\langle
\Psi_0 \left| \frac{Z\vec{r}_i}{r_i^3} \right| \Psi_m \right\rangle \right|^2 \times \\ &(E_m - E_0)^{-1} \ln |E_m - E_0|
\end{aligned}
\end{equation}
and
\begin{equation}
\begin{aligned}
\mathcal{D}^{(A)}(n,L,S) =  \sum_{m,i} \left| \left\langle
\Psi_0 \left| \frac{Z\vec{r}_i}{r_i^3} \right| \Psi_m \right\rangle \right|^2 (E_m - E_0)^{-1}  \ .
\end{aligned}
\end{equation}
The $E_m$ and $\Psi_m$ denote the intermediate energy levels and corresponding wave functions, $E_0$ and $\Psi_0$ denote the initial state.

Using the commutation relation
\begin{equation}
\begin{aligned}
(E_m-E_0)&\langle \Psi_0 | \vec{p} | \Psi_m\rangle = \\& \langle \Psi_0 |[\vec{p},H]| \Psi_m\rangle =-i Z  \langle \Psi_0 |\frac{\vec{r}}{r^3}| \Psi_m \rangle,
\end{aligned}
\end{equation}
one can get the BL in velocity-acceleration (v-a) gauge,
\begin{equation}
\begin{aligned}
\mathcal{N}^{(VA)}(n,L,S) =&  \sum_{m,i}   \left\langle
\Psi_0 \left| \frac{Z\vec{r}_i}{r_i^3} \right| \Psi_m \right\rangle    \left\langle
\Psi_0 \left|  \vec{p}_i \right| \Psi_m \right\rangle  \times \\ & \ln |E_m - E_0|
\end{aligned}
\end{equation}
and
\begin{equation}
\begin{aligned}
\mathcal{D}^{(VA)}(n,L,S) =  \sum_{m,i}   \left\langle
\Psi_0 \left| \frac{Z\vec{r}_i}{r_i^3} \right| \Psi_m \right\rangle    \left\langle
\Psi_0 \left|  \vec{p}_i \right| \Psi_m \right\rangle     \ .
\end{aligned}
\end{equation}
And the following expression of $\mathcal{D}^{(A)}$,
\begin{equation}
\begin{aligned}
\mathcal{D}^{(A)}(n,L,S) = \sum_{m,i} \left| \left\langle
\Psi_0 \left| \vec{p}_i \right| \Psi_m \right\rangle \right|^2 (E_m - E_0) \; .
\end{aligned}
\end{equation}

The last formular is similar to the expression of averaged dipole polarizability of helium if we insert $1/(E_m - E_0)^4$ in each term of the summation. Because of the absence of these factors, the high energy intermediate states also have a great contribution to $\mathcal{D}^{(A)}$. For the numerator $\mathcal{N}^{(A)}$, the additional terms $\ln|E_m-E_0|$ make the high energy intermediate levels more important. As a correction term, the numerical value of BL is small. However, it's not clear that how much the interval of intermediate energy levels is needed to accurately calculate BL of helium. Drake and Goldman pointed out that to obtain a correct convergent result of BL of helium the maximum of intermediate energy should exceed $10^6$~\cite{drake1999bethe}.

\section{RESULTS AND DISCUSSIONS}
The maximum energy of intermediate states, $E_{max}$, is crucial to the precision calculation of BL. In the case of hydrogen, Ref.~\cite{tang2013bethe-logarithm} shows that the $E_{max}$ generated by $B$-spline basis has a close relationship with the first non-zero knot, $T_1$. In the case of helium, firstly, we do the calculations in $r_{max}=200$, $\tau=0.0875$ and $r_{max}=400$, $\tau=0.0475$ respectively, letting the first non-zero knot in the range of $10^{-5}-10^{-6}$. In the two cases, under total number of $B$-splines $N=50$ and partial-wave expansion length $l_{max}=2$, the first non-zero knot and the $E_{max}$ are $T_1=2.38 \times 10^{-6}$, $E_{max}=1.85 \times 10^{15}$ and $T_1=1.17 \times 10^{-6}$, $E_{max}=7.50 \times 10^{14}$ respectively. Table~\ref{tab1} shows in $r_{max}=200$ and $r_{max}=400$, the values of BL in acceleration gauge for $1\; ^1S$ state of helium under different $N$ with fixed $l_{max}=2$. The extrapolated values of energy are $E=-2.903724(1)$ and $E=-2.903723(1)$, having $7$ significant digits compared with the benchmark result $E=-2.903 724 377 034 1195$~\cite{drake2006springer}. And the extrapolated values of BL are $4.370159\emph{6(3)}$ and $4.370160(1)$, having $7$ significant digits compared with the other high precise results. The last two digits in the first result of BL are invalid due to the limitation of the order of significant digits of the initial state energy. These calculations imply that the two sets of intermediate states generated in two boxes are all available if the initial state is calculated well enough. In the following calculations, we set $r_{max}=400$, $\tau=0.056$ and increase the total number of $B$-splines to $N=70$, partial-wave expansion length to $l_{max}=4$ to calculate the BL for $n\; ^{1}S$, $n$ up to 10, states of helium. These parameters let $T_1= 6.7 \times 10^{-8} - 3.0 \times 10^{-8}$ and $E_{max}=10^{16}-10^{18}$ in our calculations.

\subsection{Energy Levels}
Table~\ref{tab2} shows the convergence study of the ground state energy of helium as $N$ and $l_{max}$ increase. As shown in Table~\ref{tab2}, the extrapolated value for the ground stat is $E=-2.90372436(6)$, which has 9 significant digits and consistent with the benchmark result $E= -2.9037243770341195$~\cite{drake2006springer}. Table~\ref{tab3} displays the comparison of the energies for $n\; ^1S$, $n$ up to 10, states of helium. As can be seen, ours results listed in column 2 have 9 significant digits and are consistent with the values given by Ref.~\cite{drake2006springer}. These results provide the suitable initial states for the following calculations of BL.

\subsection{Bethe Logarithm}
\begin{figure}
\centering
\includegraphics[scale=.32]{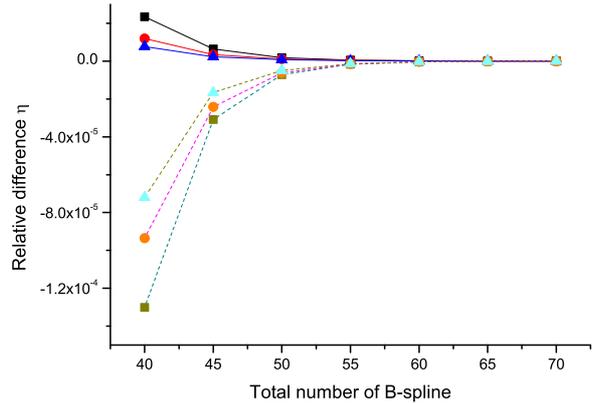}
\caption{Relative difference $\eta$ under different total number of $B$-spline $N$ and partial-wave expansion length $l_{max}$. The relative difference means $\eta(x)=(x-c)/c$, where $x$ denotes our value and $c$ denotes the other high precise result. Points in solid lines and dash lines denote values in acceleration gauge and v-a gauge respectively. Square, circle and triangle points denote the results under $l_{max}=$ 2, 3 and 4 respectively.}
\label{fig1}
\end{figure}

The convergence study of BL in two gauges for the ground state of helium are displayed in Table~\ref{tab4} and Table~\ref{tab5} respectively. From these two tables, it is can be found that the values in two gauges gradually close to the other high precise results with $N$ and $l_{max}$ increasing, as shown figure~\ref{fig1}. Our extrapolated values of BL for the ground state of helium are $\beta=4.37016022(5)$ and $\beta=4.3701601(1)$, which have $9$ and $8$ significant digits respectively. Our results are consistent with the other high precise results. And the result in acceleration gauge has 3 significant digits more precise than the result of $B$-spline basis (see Table~\ref{tab6}).

Our results of BL in two gauges for $n\; ^1S$, n up to 10, states of helium are tabulated in Table~\ref{tab6}, which also compared to the other literatures results available. In this table, our results are listed in the 2th column, the upper values in each cell are of acceleration gauge and the lower values are of v-a gauge, and so are the results of $B$-spline basis listed in the 3th column. The values in the 5th column are truncated to 10 digits. 

As shown in this table, our results in two gauges are self-consistent, having at least 7 significant digits. For $1 \; ^1S - 4 \; ^1S$ states, our results in acceleration gauge are accurate to 9 significant digits, satisfied the current experimental demand. One can find that our results of the first four states remedy the inconsistencies between the results of $B$-spline basis and the other high precise results or between the results of $B$-spline basis in two gauges. For the $4 \; ^1S$ states, our result in acceleration gauge is in the middle of the two high precise results. Except for the $5 \; ^1S$ states, our results are consistent with these high precise results. For the $8 \; ^1S - 10 \; ^1S$ states, a good agreement can be observed between our results and the results of asymptotic expansions. In our calculations, most of the knots sequence of $B$-splines are concentrated in the first half of the box. We think further optimization of the knots sequence of $B$-splines may be helpful to improve the accuracy of our results for higher lying, including $4 \; ^1S$ and $5 \; ^1S$, states.

\section{SUMMARY}
In this article, the BL for $n\; ^1S$, $n$ up to 10, states of helium in acceleration gauge and velocity-acceleration gauge are calculated by using H-$B$-spline basis set. The extrapolated results in two gauges are self-consistent and are accurate to at least 7 significant digits. In acceleration gauge, the results of BL for $1 \; ^1S - 4 \; ^1S$ states are accurate to 9 significant digits. Since the H-$B$-spline basis has capable of describing the two-electron coalescences, our results remedy the recent results of $B$-spline basis for $^1 S$ states of helium. Especial for the ground state, our result has 3 significant digits more accurate than the result of $B$-spline basis. Our results are also compared with the other high precise results. The comparison shows a very good agreement except for $5 \; ^1S$ state. For the higher lying, including $4 \; ^1S$ and $5 \; ^1S$, states, more precise results of using H-$B$-spline basis might be obtained by optimizing the knots sequence of $B$-splines. 

Furthermore, to overcome the numerical linear correlation problem, we developed the MGSEPS. It is a MPI parallel program and is not limited by the computing decimal digits. This program enables us to complete our calculations and saves us lots of time. This program, we think, will bring more possibilities to precision calculations.

\section{ACKNOWLEDGMENTS}
The authors thank Wan-Ping Zhou and Xue-Song Mei for the meaningful discussions. The numerical calculations in this article have been done on the supercomputing system in the Supercomputing Center of Wuhan University. This work is supported by the National Natural Science Foundation of China (No.11674253), (No.91536120) and  (No.11504094).

\makegapedcells
\setcellgapes{4pt}
\begin{table*}[!htbp]
\caption{Convergence of BL for $1\; ^1S$ state of helium as total number of $B$-splines $N$ increases with fixed partial-wave expansion length $l_{max}=2$ in $r_{max}=200$ and $r_{max}=400$ respectively. Numbers in parentheses are computational uncertainties. Numbers in italics are invalid. Units are a.u.}
\label{tab1}
\begin{tabular}{C{2cm}L{3.5cm}L{3.5cm}}
\hline
\hline
N&\qquad $r_{max}=200$&\qquad $r_{max}=400$\\
\hline
20&4.378 &4.37 \\
25&4.371 &4.369 \\
30&4.3703 &4.3705 \\
35&4.37019 &4.3702 \\
40&4.370167 &4.37017 \\
45&4.370161 &4.370164 \\
50&4.3701599 &4.370161 \\
$\infty$&4.370159\emph{6(3)}&4.370160(1) \\
Ref.~\cite{drake1999bethe}&4.370160218(3) \\
Ref.~\cite{PhysRevA.81.022507} &4.3701602229(1)\\
Ref.~\cite{KorobovBL} &4.3701602230703(3)\\
\hline
\hline
\end{tabular}
\end{table*}

\makegapedcells
\setcellgapes{4pt}
\begin{table*}[!htbp]
\caption{Convergence study for the ground state energy of helium as the total number of $B$-splines $N$ and the partial-wave expansion length $l_{max}$ increase. Numbers in parentheses are computational uncertainties. Units are a.u.}
\label{tab2}
\begin{tabular}{C{1cm}L{2.5cm}L{2.5cm}L{2.5cm}L{2.5cm}L{2.5cm}L{3.5cm} }
\hline
\hline
$N \backslash l_{max}$&\qquad 1&\qquad 2&\qquad 3&\qquad 4\\
\hline
40&-2.9035 &-2.90363 &-2.9036 &-2.9036 \\
45&-2.90366 &-2.90369 &-2.90370 &-2.90371 \\
50&-2.90370 &-2.90371 &-2.90371 &-2.903720 \\
55&-2.90371 &-2.903721 &-2.90372 &-2.9037231 \\
60&-2.903722 &-2.9037231 &-2.903723 &-2.9037239 \\
65&-2.9037235 &-2.9037238 &-2.9037241 &-2.9037242 \\
70&-2.9037239 &-2.9037241 &-2.9037242 &-2.903724306 \\
$\infty$&&&&-2.90372436(6)\\
\hline
\hline
\end{tabular}
\end{table*}

\makegapedcells
\setcellgapes{4pt}
\begin{table*}[!htbp]
\caption{Comparison of the energies for $n\; ^1S$, $n$ up to 10, states of helium. Numbers in parentheses of the extrapolated values are the computational uncertainties. Units are a.u.}
\label{tab3}
\begin{tabular}{C{1cm}L{3cm}L{4cm}L{3cm}L{4cm}}
\hline
\hline
n&\qquad $^{1}S$&\qquad Ref.~\cite{drake2006springer}\\
\hline
1&-2.90372436(6)&-2.9037243770341195 \\
2&-2.14597403(3)&-2.145974046054419(6) \\
3&-2.06127197(3)&-2.061271989740911(5) \\
4&-2.03358670(4)&-2.03358671703072(1) \\
5&-2.02117684(4)&-2.021176851574363(5) \\
6&-2.01456308(4)&-2.01456309844660(1) \\
7&-2.01062575(3)&-2.01062577621087(2) \\
8&-2.00809359(3)&-2.00809362210561(4) \\
9&-2.00636952(4)&-2.00636955310785(3) \\
10&-2.00514299(9)&-2.00514299174800(8) \\
\hline
\hline
\end{tabular}
\end{table*}

\makegapedcells
\setcellgapes{4pt}
\begin{table*}[!htbp]
\caption{Convergence study of BL in acceleration gauge for the ground state of helium as the total number of $B$-splines $N$ and the partial-wave expansion length $l_{max}$ increase. Numbers in parentheses are computational uncertainties. Units are a.u.}
\label{tab4}
\begin{tabular}{C{1cm}L{2.5cm}L{2.5cm}L{2.5cm}L{2.5cm}L{2.5cm}L{2.5cm}}
\hline
\hline
$N \backslash l_{max}$&\qquad 1&\qquad 2&\qquad 3&\qquad 4\\
\hline
40&4.3704 &4.3702 &4.3702 &4.37019 \\
45&4.3690 &4.37018 &4.37017 &4.370170 \\
50&4.370336 &4.370168 &4.370165 &4.370163 \\
55&4.370330 &4.370162 &4.370161 &4.3701613 \\
60&4.370328 &4.370161 &4.3701608 &4.3701606 \\
65&4.3703275 &4.3701606 &4.3701604 &4.3701603 \\
70&4.3703272 &4.3701604 &4.3701603 &4.37016027 \\
$\infty$& & & & 4.37016022(5)\\
\hline
\hline
\end{tabular}
\end{table*}

\makegapedcells
\setcellgapes{4pt}
\begin{table*}[!htbp]
\caption{Convergence study of BL in velocity-acceleration gauge for the ground state of helium as the total number of $B$-splines $N$ and the partial-wave expansion length $l_{max}$ increase. Numbers in parentheses are computational uncertainties. Units are a.u.}
\label{tab5}
\begin{tabular}{C{1cm}L{2.5cm}L{2.5cm}L{2.5cm}L{2.5cm}L{2.5cm}}
\hline
\hline
$N \backslash l_{max}$&\qquad 1&\qquad 2&\qquad 3&\qquad 4\\
\hline
40&4.369 &4.3695 &4.3697 &4.3698 \\
45&4.368 &4.37002 &4.37005 &4.37008 \\
50&4.37015 &4.37012 &4.37013 &4.370138 \\
55&4.370164 &4.370152 &4.370153 &4.370154 \\
60&4.370167 &4.370158 &4.3701584 &4.3701585 \\
65&4.3701682 &4.370159 &4.3701596 &4.3701597 \\
70&4.3701684 &4.37016006 &4.37016002 &4.37016004 \\

$\infty$& & & & 4.3701601(1)\\
\hline
\hline
\end{tabular}
\end{table*}

\makegapedcells
\setcellgapes{4pt}
\begin{table*}[!htbp]
\caption{Comparison of BL for $n\; ^{1}S$, $n$ up to 10, states of helium. The upper values in each cell of the 2th and 3th column are of acceleration gauge and the lower values are of velocity-acceleration gauge respectively. Numbers in parentheses are computational uncertainties. Units are a.u.}
\label{tab6}
\begin{tabular}{C{1.5cm}L{3.5cm}L{3cm}L{4cm}L{3.5cm}}
\hline
\hline
States&\quad this work &\quad $B$-spline~\cite{zhang2019calculations}&\qquad references &asymptotic expansions\\
\hline
$1\; ^{1}S$&4.37016022(5)&4.37034(2)  &4.370160218(3)$^{\rm a}$      &\\
		   &4.3701601(1) &4.37014(2)  &4.3701602229(1)$^{\rm b}$     &\\
		   &             &            &4.3701602230703(3)$^{\rm c}$  &\\
$2\; ^{1}S$&4.36641271(1)&4.36643(1)  &4.36641272(7)$^{\rm a}$       &4.366412729$^{\rm d}$\\
           &4.3664127(1)&4.366412(1)  &4.3664127262(1)$^{\rm b}$     &4.366378229$^{\rm c}$\\
           &             &            &4.366412726417(1)$^{\rm c}$   &\\
$3\; ^{1}S$&4.36916480(6)&4.369170(1) &4.369164871(8)$^{\rm a}$      &4.369164888$^{\rm d}$\\
		   &4.3691648(1) &4.3691643(2)&4.369164860824(2)$^{\rm c}$   &4.369164809$^{\rm c}$\\
$4\; ^{1}S$&4.36989065(5)&4.369893(1) &4.36989066(1)$^{\rm a}$       &4.369890657$^{\rm d}$\\
		   &4.3698906(1) &4.3698903(5)&4.36989063236(1)$^{\rm c}$    &4.369890661$^{\rm c}$\\
$5\; ^{1}S$&4.3701520(1) &4.370152(3) &4.3701517(1)$^{\rm a}$        &4.370152093$^{\rm d}$\\
		   &4.3701519(1) &4.3701511(2)&4.37015179631(1)$^{\rm c}$    &4.370151761$^{\rm c}$\\
$6\; ^{1}S$&4.370267(1)  &4.37027(1)  &4.37026697432(3)$^{\rm c}$    &4.370267364$^{\rm d}$\\
		   &4.370267(1)  &4.370266(2) &                              &4.370266961$^{\rm c}$\\
$7\; ^{1}S$&4.370326(1)  &4.37033(1)  &4.37032526176(2)$^{\rm c}$    &4.370325649$^{\rm d}$\\
		   &4.370326(1)  &4.37033(1)  &                              &4.370325274$^{\rm c}$\\
$8\; ^{1}S$&4.370359(2)  &4.37034(4)  & 				             &4.370358160$^{\rm d}$\\
		   &4.370359(2)  &4.37034(2)  &                              &4.370357839$^{\rm c}$\\
$9\; ^{1}S$&4.370378(2)  &            & 				             &4.370377682$^{\rm d}$\\
		   &4.370378(2)  &            &                              &4.370377414$^{\rm c}$\\
$10\; ^{1}S$&4.370389(1) &            & 				             &4.370390095$^{\rm d}$\\
		   &4.370388(1)  &            &                              &4.370389875$^{\rm c}$\\
\hline
\hline
\end{tabular}
\begin{tablenotes}
	\footnotesize
	\item[1] $^{\rm a}$ Drake and Goldman~\cite{drake1999bethe}
	\item[2] $^{\rm b}$ Yerokhin and Pachucki~\cite{PhysRevA.81.022507}
	\item[3] $^{\rm c}$ Korobov~\cite{KorobovBL}
	\item[4] $^{\rm d}$ Drake~\cite{drake2001qed}
\end{tablenotes}
\end{table*}

\bibliography{Bethe_Logarithm}

\begin{thebibliography}{32}
\expandafter\ifx\csname natexlab\endcsname\relax\def\natexlab#1{#1}\fi
\expandafter\ifx\csname bibnamefont\endcsname\relax
  \def\bibnamefont#1{#1}\fi
\expandafter\ifx\csname bibfnamefont\endcsname\relax
  \def\bibfnamefont#1{#1}\fi
\expandafter\ifx\csname citenamefont\endcsname\relax
  \def\citenamefont#1{#1}\fi
\expandafter\ifx\csname url\endcsname\relax
  \def\url#1{\texttt{#1}}\fi
\expandafter\ifx\csname urlprefix\endcsname\relax\def\urlprefix{URL }\fi
\providecommand{\bibinfo}[2]{#2}
\providecommand{\eprint}[2][]{\url{#2}}

\bibitem[{\citenamefont{Pachucki et~al.}(2017)\citenamefont{Pachucki,
  Patk{\'o}{\v{s}}, and Yerokhin}}]{pachucki2017testing}
\bibinfo{author}{\bibfnamefont{K.}~\bibnamefont{Pachucki}},
  \bibinfo{author}{\bibfnamefont{V.}~\bibnamefont{Patk{\'o}{\v{s}}}},
  \bibnamefont{and} \bibinfo{author}{\bibfnamefont{V.~A.}
  \bibnamefont{Yerokhin}}, \bibinfo{journal}{Physical Review A}
  \textbf{\bibinfo{volume}{95}}, \bibinfo{pages}{062510}
  (\bibinfo{year}{2017}).

\bibitem[{\citenamefont{Pastor et~al.}(2004)\citenamefont{Pastor, Giusfredi,
  Natale, Hagel, de~Mauro, and Inguscio}}]{PhysRevLett.92.023001}
\bibinfo{author}{\bibfnamefont{P.~C.} \bibnamefont{Pastor}},
  \bibinfo{author}{\bibfnamefont{G.}~\bibnamefont{Giusfredi}},
  \bibinfo{author}{\bibfnamefont{P.~D.} \bibnamefont{Natale}},
  \bibinfo{author}{\bibfnamefont{G.}~\bibnamefont{Hagel}},
  \bibinfo{author}{\bibfnamefont{C.}~\bibnamefont{de~Mauro}}, \bibnamefont{and}
  \bibinfo{author}{\bibfnamefont{M.}~\bibnamefont{Inguscio}},
  \bibinfo{journal}{Phys. Rev. Lett.} \textbf{\bibinfo{volume}{92}},
  \bibinfo{pages}{023001} (\bibinfo{year}{2004}).

\bibitem[{\citenamefont{Zheng and Sun}(2017)}]{zheng2017x}
\bibinfo{author}{\bibfnamefont{X.}~\bibnamefont{Zheng}} \bibnamefont{and}
  \bibinfo{author}{\bibfnamefont{Y.}~\bibnamefont{Sun}},
  \bibinfo{journal}{Phys. Rev. Lett.} \textbf{\bibinfo{volume}{118}},
  \bibinfo{pages}{063001} (\bibinfo{year}{2017}).

\bibitem[{\citenamefont{Safronova et~al.}(2018)\citenamefont{Safronova, Budker,
  DeMille, Kimball, Derevianko, and Clark}}]{RevModPhys.90.025008}
\bibinfo{author}{\bibfnamefont{M.~S.} \bibnamefont{Safronova}},
  \bibinfo{author}{\bibfnamefont{D.}~\bibnamefont{Budker}},
  \bibinfo{author}{\bibfnamefont{D.}~\bibnamefont{DeMille}},
  \bibinfo{author}{\bibfnamefont{D.~F.~J.} \bibnamefont{Kimball}},
  \bibinfo{author}{\bibfnamefont{A.}~\bibnamefont{Derevianko}},
  \bibnamefont{and} \bibinfo{author}{\bibfnamefont{C.~W.} \bibnamefont{Clark}},
  \bibinfo{journal}{Rev. Mod. Phys.} \textbf{\bibinfo{volume}{90}},
  \bibinfo{pages}{025008} (\bibinfo{year}{2018}).

\bibitem[{\citenamefont{Schwartz}(1964)}]{schwartz1964fine}
\bibinfo{author}{\bibfnamefont{C.}~\bibnamefont{Schwartz}},
  \bibinfo{journal}{Physical Review} \textbf{\bibinfo{volume}{134}}
  (\bibinfo{year}{1964}).

\bibitem[{\citenamefont{Wu et~al.}(2018)\citenamefont{Wu, Yang, Zhang, Zhang,
  Qiao, Shi, and Tang}}]{PhysRevA.98.040501}
\bibinfo{author}{\bibfnamefont{F.-F.} \bibnamefont{Wu}},
  \bibinfo{author}{\bibfnamefont{S.-J.} \bibnamefont{Yang}},
  \bibinfo{author}{\bibfnamefont{Y.-H.} \bibnamefont{Zhang}},
  \bibinfo{author}{\bibfnamefont{J.-Y.} \bibnamefont{Zhang}},
  \bibinfo{author}{\bibfnamefont{H.-X.} \bibnamefont{Qiao}},
  \bibinfo{author}{\bibfnamefont{T.-Y.} \bibnamefont{Shi}}, \bibnamefont{and}
  \bibinfo{author}{\bibfnamefont{L.-Y.} \bibnamefont{Tang}},
  \bibinfo{journal}{Phys. Rev. A} \textbf{\bibinfo{volume}{98}},
  \bibinfo{pages}{040501} (\bibinfo{year}{2018}).

\bibitem[{\citenamefont{van Leeuwen and Vassen}(2006)}]{Leeuwen_2006}
\bibinfo{author}{\bibfnamefont{K.~A.~H.} \bibnamefont{van Leeuwen}}
  \bibnamefont{and} \bibinfo{author}{\bibfnamefont{W.}~\bibnamefont{Vassen}},
  \bibinfo{journal}{Europhysics Letters ({EPL})} \textbf{\bibinfo{volume}{76}},
  \bibinfo{pages}{409} (\bibinfo{year}{2006}).

\bibitem[{\citenamefont{Shiner et~al.}(1995)\citenamefont{Shiner, Dixson, and
  Vedantham}}]{PhysRevLett.74.3553}
\bibinfo{author}{\bibfnamefont{D.}~\bibnamefont{Shiner}},
  \bibinfo{author}{\bibfnamefont{R.}~\bibnamefont{Dixson}}, \bibnamefont{and}
  \bibinfo{author}{\bibfnamefont{V.}~\bibnamefont{Vedantham}},
  \bibinfo{journal}{Phys. Rev. Lett.} \textbf{\bibinfo{volume}{74}},
  \bibinfo{pages}{3553} (\bibinfo{year}{1995}).

\bibitem[{\citenamefont{Morton et~al.}(2006)\citenamefont{Morton, Wu, and
  Drake}}]{PhysRevA.73.034502}
\bibinfo{author}{\bibfnamefont{D.~C.} \bibnamefont{Morton}},
  \bibinfo{author}{\bibfnamefont{Q.}~\bibnamefont{Wu}}, \bibnamefont{and}
  \bibinfo{author}{\bibfnamefont{G.~W.~F.} \bibnamefont{Drake}},
  \bibinfo{journal}{Phys. Rev. A} \textbf{\bibinfo{volume}{73}},
  \bibinfo{pages}{034502} (\bibinfo{year}{2006}).

\bibitem[{\citenamefont{Cancio~Pastor et~al.}(2012)\citenamefont{Cancio~Pastor,
  Consolino, Giusfredi, De~Natale, Inguscio, Yerokhin, and
  Pachucki}}]{PhysRevLett.108.143001}
\bibinfo{author}{\bibfnamefont{P.}~\bibnamefont{Cancio~Pastor}},
  \bibinfo{author}{\bibfnamefont{L.}~\bibnamefont{Consolino}},
  \bibinfo{author}{\bibfnamefont{G.}~\bibnamefont{Giusfredi}},
  \bibinfo{author}{\bibfnamefont{P.}~\bibnamefont{De~Natale}},
  \bibinfo{author}{\bibfnamefont{M.}~\bibnamefont{Inguscio}},
  \bibinfo{author}{\bibfnamefont{V.~A.} \bibnamefont{Yerokhin}},
  \bibnamefont{and} \bibinfo{author}{\bibfnamefont{K.}~\bibnamefont{Pachucki}},
  \bibinfo{journal}{Phys. Rev. Lett.} \textbf{\bibinfo{volume}{108}},
  \bibinfo{pages}{143001} (\bibinfo{year}{2012}).

\bibitem[{\citenamefont{Drake}(1978)}]{drake1978angular}
\bibinfo{author}{\bibfnamefont{G.~W.~F.} \bibnamefont{Drake}},
  \bibinfo{journal}{Physical Review A} \textbf{\bibinfo{volume}{18}},
  \bibinfo{pages}{820} (\bibinfo{year}{1978}).

\bibitem[{\citenamefont{Yan and Drake}(1994)}]{yan1994general}
\bibinfo{author}{\bibfnamefont{Z.}~\bibnamefont{Yan}} \bibnamefont{and}
  \bibinfo{author}{\bibfnamefont{G.~W.~F.} \bibnamefont{Drake}},
  \bibinfo{journal}{Canadian Journal of Physics} \textbf{\bibinfo{volume}{72}},
  \bibinfo{pages}{822} (\bibinfo{year}{1994}).

\bibitem[{\citenamefont{Frolov}(2000)}]{frolov2000high-precision}
\bibinfo{author}{\bibfnamefont{A.~M.} \bibnamefont{Frolov}},
  \bibinfo{journal}{Physical Review E} \textbf{\bibinfo{volume}{62}},
  \bibinfo{pages}{8740} (\bibinfo{year}{2000}).

\bibitem[{\citenamefont{Caswell and Lepage}(1986)}]{caswell1986effective}
\bibinfo{author}{\bibfnamefont{W.~E.} \bibnamefont{Caswell}} \bibnamefont{and}
  \bibinfo{author}{\bibfnamefont{G.~P.} \bibnamefont{Lepage}},
  \bibinfo{journal}{Physics Letters B} \textbf{\bibinfo{volume}{167}},
  \bibinfo{pages}{437} (\bibinfo{year}{1986}).

\bibitem[{\citenamefont{Jentschura et~al.}(2005)\citenamefont{Jentschura,
  Czarnecki, and Pachucki}}]{jentschura2005nonrelativistic}
\bibinfo{author}{\bibfnamefont{U.~D.} \bibnamefont{Jentschura}},
  \bibinfo{author}{\bibfnamefont{A.}~\bibnamefont{Czarnecki}},
  \bibnamefont{and} \bibinfo{author}{\bibfnamefont{K.}~\bibnamefont{Pachucki}},
  \bibinfo{journal}{Physical Review A} \textbf{\bibinfo{volume}{72}},
  \bibinfo{pages}{062102} (\bibinfo{year}{2005}).

\bibitem[{\citenamefont{Pachucki and Yerokhin}(2010)}]{pachucki2010fine}
\bibinfo{author}{\bibfnamefont{K.}~\bibnamefont{Pachucki}} \bibnamefont{and}
  \bibinfo{author}{\bibfnamefont{V.~A.} \bibnamefont{Yerokhin}},
  \bibinfo{journal}{Physical review letters} \textbf{\bibinfo{volume}{104}},
  \bibinfo{pages}{070403} (\bibinfo{year}{2010}).

\bibitem[{\citenamefont{Drake and Goldman}(1999)}]{drake1999bethe}
\bibinfo{author}{\bibfnamefont{G.~W.~F.} \bibnamefont{Drake}} \bibnamefont{and}
  \bibinfo{author}{\bibfnamefont{S.~P.} \bibnamefont{Goldman}},
  \bibinfo{journal}{Canadian Journal of Physics} \textbf{\bibinfo{volume}{77}},
  \bibinfo{pages}{835} (\bibinfo{year}{1999}).

\bibitem[{\citenamefont{Goldman and
  Drake}(2000{\natexlab{a}})}]{goldman2000high}
\bibinfo{author}{\bibfnamefont{S.~P.} \bibnamefont{Goldman}} \bibnamefont{and}
  \bibinfo{author}{\bibfnamefont{G.~W.~F.} \bibnamefont{Drake}},
  \bibinfo{journal}{Physical Review A} \textbf{\bibinfo{volume}{61}},
  \bibinfo{pages}{052513} (\bibinfo{year}{2000}{\natexlab{a}}).

\bibitem[{\citenamefont{Drake}(2001)}]{drake2001qed}
\bibinfo{author}{\bibfnamefont{G.~W.~F.} \bibnamefont{Drake}},
  \bibinfo{journal}{Physica Scripta} \textbf{\bibinfo{volume}{2001}},
  \bibinfo{pages}{22} (\bibinfo{year}{2001}).

\bibitem[{\citenamefont{Goldman and
  Drake}(2000{\natexlab{b}})}]{PhysRevA.61.052513}
\bibinfo{author}{\bibfnamefont{S.~P.} \bibnamefont{Goldman}} \bibnamefont{and}
  \bibinfo{author}{\bibfnamefont{G.~W.~F.} \bibnamefont{Drake}},
  \bibinfo{journal}{Phys. Rev. A} \textbf{\bibinfo{volume}{61}},
  \bibinfo{pages}{052513} (\bibinfo{year}{2000}{\natexlab{b}}).

\bibitem[{\citenamefont{Korobov}(2004)}]{PhysRevA.69.054501}
\bibinfo{author}{\bibfnamefont{V.~I.} \bibnamefont{Korobov}},
  \bibinfo{journal}{Phys. Rev. A} \textbf{\bibinfo{volume}{69}},
  \bibinfo{pages}{054501} (\bibinfo{year}{2004}).

\bibitem[{\citenamefont{Tang et~al.}(2013)\citenamefont{Tang, Zhong, Li, Qiao,
  and Shi}}]{tang2013bethe-logarithm}
\bibinfo{author}{\bibfnamefont{Y.}~\bibnamefont{Tang}},
  \bibinfo{author}{\bibfnamefont{Z.}~\bibnamefont{Zhong}},
  \bibinfo{author}{\bibfnamefont{C.}~\bibnamefont{Li}},
  \bibinfo{author}{\bibfnamefont{H.}~\bibnamefont{Qiao}}, \bibnamefont{and}
  \bibinfo{author}{\bibfnamefont{T.}~\bibnamefont{Shi}},
  \bibinfo{journal}{Physical Review A} \textbf{\bibinfo{volume}{87}}
  (\bibinfo{year}{2013}).

\bibitem[{\citenamefont{Zhong et~al.}(2013)\citenamefont{Zhong, Yan, and
  Shi}}]{PhysRevA.88.052520}
\bibinfo{author}{\bibfnamefont{Z.-X.} \bibnamefont{Zhong}},
  \bibinfo{author}{\bibfnamefont{Z.-C.} \bibnamefont{Yan}}, \bibnamefont{and}
  \bibinfo{author}{\bibfnamefont{T.-Y.} \bibnamefont{Shi}},
  \bibinfo{journal}{Phys. Rev. A} \textbf{\bibinfo{volume}{88}},
  \bibinfo{pages}{052520} (\bibinfo{year}{2013}).

\bibitem[{\citenamefont{Schwartz}(1961)}]{PhysRev.123.1700}
\bibinfo{author}{\bibfnamefont{C.}~\bibnamefont{Schwartz}},
  \bibinfo{journal}{Phys. Rev.} \textbf{\bibinfo{volume}{123}},
  \bibinfo{pages}{1700} (\bibinfo{year}{1961}).

\bibitem[{\citenamefont{Korobov}(2012)}]{PhysRevA.85.042514}
\bibinfo{author}{\bibfnamefont{V.~I.} \bibnamefont{Korobov}},
  \bibinfo{journal}{Phys. Rev. A} \textbf{\bibinfo{volume}{85}},
  \bibinfo{pages}{042514} (\bibinfo{year}{2012}).

\bibitem[{\citenamefont{Korobov}(2019)}]{KorobovBL}
\bibinfo{author}{\bibfnamefont{V.~I.} \bibnamefont{Korobov}},
  \bibinfo{journal}{Phys. Rev. A} \textbf{\bibinfo{volume}{100}},
  \bibinfo{pages}{012517} (\bibinfo{year}{2019}).

\bibitem[{\citenamefont{Yang et~al.}(2017)\citenamefont{Yang, Mei, Shi, and
  Qiao}}]{yang2017application}
\bibinfo{author}{\bibfnamefont{S.-J.} \bibnamefont{Yang}},
  \bibinfo{author}{\bibfnamefont{X.-S.} \bibnamefont{Mei}},
  \bibinfo{author}{\bibfnamefont{T.-Y.} \bibnamefont{Shi}}, \bibnamefont{and}
  \bibinfo{author}{\bibfnamefont{H.-X.} \bibnamefont{Qiao}},
  \bibinfo{journal}{Physical Review A} \textbf{\bibinfo{volume}{95}},
  \bibinfo{pages}{062505} (\bibinfo{year}{2017}).

\bibitem[{\citenamefont{Bailey et~al.}(2002)\citenamefont{Bailey, Yozo, Li, and
  Thompson}}]{bailey2002arprec}
\bibinfo{author}{\bibfnamefont{D.~H.} \bibnamefont{Bailey}},
  \bibinfo{author}{\bibfnamefont{H.}~\bibnamefont{Yozo}},
  \bibinfo{author}{\bibfnamefont{X.~S.} \bibnamefont{Li}}, \bibnamefont{and}
  \bibinfo{author}{\bibfnamefont{B.}~\bibnamefont{Thompson}},
  \bibinfo{type}{Tech. Rep.}, \bibinfo{institution}{Lawrence Berkeley National
  Lab.(LBNL), Berkeley, CA (United States)} (\bibinfo{year}{2002}).

\bibitem[{\citenamefont{Zhang et~al.}(2019)\citenamefont{Zhang, Shen, Xiao,
  Zhang, and Shi}}]{zhang2019calculations}
\bibinfo{author}{\bibfnamefont{Y.-H.} \bibnamefont{Zhang}},
  \bibinfo{author}{\bibfnamefont{L.-J.} \bibnamefont{Shen}},
  \bibinfo{author}{\bibfnamefont{C.-M.} \bibnamefont{Xiao}},
  \bibinfo{author}{\bibfnamefont{J.-Y.} \bibnamefont{Zhang}}, \bibnamefont{and}
  \bibinfo{author}{\bibfnamefont{T.-Y.} \bibnamefont{Shi}},
  \bibinfo{journal}{arXiv preprint arXiv:1903.08802}  (\bibinfo{year}{2019}).

\bibitem[{\citenamefont{De~Boor}(1978)}]{deboor1978a}
\bibinfo{author}{\bibfnamefont{C.}~\bibnamefont{De~Boor}},
  \bibinfo{journal}{Mathematics of Computation} \textbf{\bibinfo{volume}{34}},
  \bibinfo{pages}{325} (\bibinfo{year}{1978}).

\bibitem[{\citenamefont{Drake}(2006)}]{drake2006springer}
\bibinfo{author}{\bibfnamefont{G.~W.} \bibnamefont{Drake}},
  \emph{\bibinfo{title}{Springer handbook of atomic, molecular, and optical
  physics}} (\bibinfo{publisher}{Springer Science \& Business Media},
  \bibinfo{year}{2006}).

\bibitem[{\citenamefont{Yerokhin and Pachucki}(2010)}]{PhysRevA.81.022507}
\bibinfo{author}{\bibfnamefont{V.~A.} \bibnamefont{Yerokhin}} \bibnamefont{and}
  \bibinfo{author}{\bibfnamefont{K.}~\bibnamefont{Pachucki}},
  \bibinfo{journal}{Phys. Rev. A} \textbf{\bibinfo{volume}{81}},
  \bibinfo{pages}{022507} (\bibinfo{year}{2010}).

\end{thebibliography}

\end{document}